\documentclass[]{spie}  %>>> use for US letter paper
\usepackage[]{graphicx}
\usepackage[]{xspace}
\usepackage{amssymb}
\usepackage{textcomp}
\usepackage{mathcomp}
\usepackage{rotating}

\newcommand{\NACO}{N{\large \bf A}C{\large \bf O}}

\newcommand{\naco}{N{\scriptsize A}C{\scriptsize O}\xspace}
\newcommand{\arcsec}{\mbox{$''$}}

\def\bc{\begin{center}}
\def\ec{\end{center}}
\def\bl{\begin{list}{$\bullet$}{}}
\def\el{\end{list}}

\def\fprop{\ensuremath{\gamma}}
\def\fprop1{\ensuremath{\gamma^{-1}}}

\def\facprop{$\gamma$\xspace}
\def\facprop1{$\gamma^{-1}$\xspace}

\def\mas{{\sl mas}\xspace}

\def\pass1{{\sl PASS-I}\xspace}
\def\pass2{{\sl PASS-II}\xspace}

\def\bc{\begin{center}}
\def\ec{\end{center}}

\def\benum{\begin{enumerate}}
\def\eenum{\end{enumerate}}

\def\bv{\begin{verbatim}}
\def\endv{\end{verbatim}}

\def\be{\begin{equation}}
\def\ee{\end{equation}}

\def\bes{\begin{equation*}}
\def\ees{\end{equation*}}

\def\bl{\begin{list}{$\bullet$}{}} %gÈnÈral
\def\el{\end{list}}

\def\bi{\begin{list}{$\bullet$}{}} %gÈnÈral
\def\ei{\end{list}}

\def\bls{\begin{list}} %special
\def\els{\end{list}}
      
\def\bt{\begin{table}}
\def\et{\end{table}}

\def\bsplit{\begin{split}} %ne marche pas...
\def\esplit{\end{split}}

\def\btm{\begin{MAJ}} %trait dans la marge
\def\etm{\end{MAJ}}

\def\arcsec{\hbox{$^{\prime\prime}$}}
\def\phi{\varphi}

\title{Image quality and high contrast  improvements on VLT/\NACO}

\author{Julien H. V. Girard\supit{a}, 
Jared O'Neal\supit{a}, 
Dimitri Mawet\supit{a,b}, 
Markus Kasper\supit{c}, 
G\'erard Zins\supit{d},\\
Beno\^{i}t Neichel\supit{e},
Johann Kolb\supit{c},
Valentin Christiaens\supit{f,a}
and
Martin Tourneboeuf\supit{g}
\skiplinehalf
\supit{a}European Southern Observatory (ESO), Casilla 19001, Vitacura, Santiago, Chile\\
\supit{b}Jet Propulsion Laboratory, California Institute of Technology, Pasadena, CA 91109, USA\\
\supit{c}European Southern Observatory, Karl-Schwarzschild-Stra\ss e 2, D-85748 Garching, Germany\\
\supit{d}IPAG, Universit\'e Joseph Fourier / CNRS, BP 53, F-38041 Grenoble, France\\
\supit{e}Gemini Observatory, AURA/Casilla 603, La Serena, Chile\\
\supit{f}D\'{e}partement d'Astrophysique, Universit\'{e} de Li\`{e}ge, 17 all\'{e}e du Six Ao\^{u}t, 4000, Li\`{e}ge, Belgium\\
\supit{g} Centre for Astro-Engineering, P. Universidad Cat\'{o}lica, Vicu\~{n}a Mackena 4860, Santiago, Chile}

\authorinfo{Corresponding author's e-mail: jgirard@eso.org}
\begin{document} 
\maketitle

%%%%%%%%%%%%%%%%%%%%%%%%%%%%%%%%%%%%%%%%%%%%%%%%%%%%%%%%%%%%% 
\begin{abstract}

 \naco\footnote{\naco: Nasmyth Adaptive Optics System {\bf NA}OS and the Near-Infrared Imager and Spectrograph {\bf CO}NICA} is the famous and versatile diffraction limited NIR imager and spectrograph with which ESO celebrated 10 years of Adaptive Optics at the VLT. Since two years a substantial effort has been put in to understanding and fixing issues that directly affect the image quality and the high contrast performances of the instrument. Experiments to compensate the non-common-path aberrations and recover the highest possible Strehl ratios have been carried out successfully and a plan is hereafter described to perform such measurements regularly. The drift associated to pupil tracking since 2007 was fixed in October 2011. \naco is therefore even better suited for high contrast imaging and can be used with coronagraphic masks in the image plane. Some contrast measurements are shown and discussed. The work accomplished on \naco will serve as reference for the next generation instruments on the VLT, especially those working at the diffraction limit and making use of angular differential imaging (i.e.~SPHERE\cite{beuzit2010lyot}, VISIR\cite{kerber2010spie}, and possibly ERIS\cite{marchetti2012spie}).

%{\tt spie.cls} (Version 3.3).  
\end{abstract}

%>>>> Include a list of keywords after the abstract 

\keywords{Adaptive Optics, High Angular Resolution, High Contrast Imaging, Very Large Telescope, Infrared Astronomy, Phase Diversity, Strehl Ratio, Calibration}

%%%%%%%%%%%%%%%%%%%%%%%%%%%%%%%%%%%%%%%%%%%%%%%%%%%%%%%%%%%%%
\section{\naco: still in operation}
\label{sec:intro}  % \label{} allows reference to this section

\naco \cite{lenzen2003,rousset2003} is the adaptive optics (AO) fed near infrared (NIR) imager and spectrometer at the 8-meter Very Large Telescope (VLT) ran by the European Southern Observatory (ESO). It was commissioned in 2001\protect\cite{brandner2002} and offered to the community in october 2002  (period 70). Refer to Lenzen et al. \cite{lenzen2003} for a description of the scientific instrument CONICA and its many operating modes, and Rousset et al. \cite{rousset2003} for an overview of the NAOS AO system.

In our 2010 paper \cite{girard2010spie} we gave an update on \naco and its new operational modes. At the time, it was thought that \naco would be decommissioned within a year or two, and so the paper emphasized the instrument scientific impact and originality. The community however strongly expressed its desire to keep \naco longer. ESO's Scientific Technical Committee (STC) recently decided to keep \naco operational until at least 2014 with a possible change of UT4 nasmyth focus (depending on MUSE's arrival at Paranal). Indeed, no other instruments can currently replace \naco for the precise follow-up of the very red galactic center region and it also provides unique L-band imaging capabilities that are very useful for exoplanet imaging and characterization (astrometry, photometry, etc.).

Among the ``ESO top 10 discoveries'' advertised on the ESO official website, four were achieved fully or partly with \naco:
\begin{enumerate}
\item  Stars orbiting the Milky Way black hole\cite{schoedel2002, gillessen2009}
\setcounter{enumi}{2}
\item  First image of an exoplanet\cite{chauvin2004}
\setcounter{enumi}{6}
\item  Flares from the supermassive black hole at the centre of the Milky Way\cite{eisenhauer2005, eckart2008}
\setcounter{enumi}{7}
\item  Direct measurements of the spectra of exoplanets and their atmospheres \cite{janson2010}
\end{enumerate}

\naco's portfolio of modes are so numerous that there is not enough time to exploit them all and they therefore compete with each other. The idea now is thus to keep only the pertinent operational modes, ``freezing'' and ``tuning '' \naco for better and more efficient (less technical downtime and resources) operations with the same or better science return and impact. By going through this process, the Instrumentation and Operations Team (IOT) is learning a lot and building AO awareness and competence to prepare for the arrival of the next-generation AO instruments.

\section{\naco, a high contrast instrument \& testbed} 
\label{sec:nacohc}

The high contrast field, both from the ground and from space, is very active and evolving rapidly as reviewed by Mawet\cite{mawet2012spie} and Kasper\cite{kasper2012spie} in this conference. Thanks to the coming of extreme-adaptive optics facilities at major observatories (SPHERE at the VLT, GPI at Gemini, P3K at Palomar, HiCIAO at Subaru). Extreme AO (ExAO)  will provide excellent image quality (Strehl ratios$\sim$90\%) to 8-m class telescopes, which will allow the use of new-generation coronagraphic systems. The baseline coronagraphs for these systems are representative of the four big families of coronagraphs: image-plane masks acting on either amplitude or phase, and pupil-plane devices acting also on either amplitude or phase, or a clever mix of all the above. At the same time, observing strategies and associated internal calibration and data reduction techniques have dramatically improved to allow for a better subtraction of the residual systematics affecting high contrast images (speckles) that are mostly due to the presence of slowly-varying optical aberrations within the instruments (static aberrations).

\naco is the current high contrast instrument at the VLT. It was also designed to be a very versatile, multi-purpose facility. Being a high-impact (section \ref{sec:intro}) and high-pressure instrument ($\sim$ 400 refereed papers \cite{telbib} in $\sim$ 11 years of operations from 2002 to 2012), the time, staff and resources allocated for non core-operations (tests, improvements, upgrades and even characterization) is limited and sparse. Nevertheless, \naco is currently the only AO system on Paranal operating with Shack-Hartmann (SH) wavefront sensors (WFS). Knowledge and experience derived from its operation is therefore very relevant to many of the future systems (AOF/GRAAL\cite{paufique2010spie} for Hawk-I, AOF/GALACSI\cite{stuik2006} for Muse, SAXO\cite{sauvage2010spie} for SPHERE, etc.) that will also use high speed SH WFS with a common SPARTA\cite{fedrigo2010spie,suarezvalles2012spie} real time computer (RTC) platform.

Though instruments commissioned and operated at the ESO Paranal Observatory are usually planned to be upgraded only on exceptional occasions (i.e.~new detector, new gas cell, etc.), \naco underwent numerous interventions and upgrades, especially to enhance its high contrast imaging capabilities opening new parameter spaces and increasing the chance to find planetary mass companions around nearby young stars. Here we present a non-exhaustive chronological list of high contrast upgrades performed on \cite{kasper2005}:

\begin{itemize}
\item 2001: Nov: \naco First-light\cite{brandner2002}! \naco includes Lyot masks (opaque and semi-transparent).
\item 2003: Simultaneous differential Imaging (SDI)mode\cite{biller2004spie} 
\item 2004: Aladdin III Detector upgrade: better cosmetics and dynamic/linearity. 
\item 2004: Four quadrant phase mask (4QPM) coronagraphs\cite{boccaletti2004} in H and Ks bands and SDI+4
\item 2004: Superachromatic half wave plate (HWP) for polarimetry: enhanced polarimetric differential imaging (PDI).
\item 2003: Low-resolution prism (R=50-400): possibly higher contrast spectroscopy and/or simultaneous spectro-photometry from J to M band.
\item 2005: CONICA cube mode.
\item 2008: Sparse Aperture Masking (SAM)\cite{tuthill2010spie}+ Pupil-tracking (PT) upgrade\cite{kasper2009}
\item 2010: Apodized Phase Plate (APP)\cite{codona2006, quanz2010, quanz2011, kenworthy2010spie}
\item 2011: APP-Spec\cite{girard2010lyot}, SAM hopping strategy\cite{lacour2011}
\item 2011: Non-Common Path Aberrations (NCPA) / Point Spread Function (PSF) full calibration.  
\item 2012:  Annular Grove Phase Mask (AGPM)\cite{mawet2005} ``Vector Vortex coronagraph'' at 4$\mu m$, SAM annular mask (to be confirmed).
\item 2013: new life on another focus? Cross-section and synergy with SPHERE? (Not yet decided).
\end{itemize}

\section{NACO Image Quality} 
\label{sec:iq}

If the NCPAs are not perfectly compensated (NAOS and/or CONICA), quasi-static speckles grow in size and number in the PSF, degrading the Strehl ratio. As they are systematics that are difficult to calibrate out, they reduce sensitivity (detection limits). Indeed, in the framework of exoplanet search, they can not easily be distinguished from faint companions in long-exposure images. Their fluctuation timescale depends on both environmental changes (i.e. temperature gradients) and instrument gravity vector (NAOS-CONICA rotates at the Nasmyth focus). Although smart observing strategies such as {\sl angular differential imaging} (ADI\cite{marois2006}), and associated post-processing techniques\cite{lafreniere2007, soummer2012} treat the residual speckle noise more and more efficiently, the prevention of this adverse effect of NCPAs before it even materializes should always be the number one priority to reach high contrasts.

The original, thorough CONICA calibrations were performed shortly after commissioning\cite{rousset2003} and sparsely by AO experts for NAOS (reference slopes, interaction matrices). The procedure used dealt separately with the NAOS and CONICA NCPAs\cite{blanc2003, hartung2003pd} and furthermore in an element-by-element approach. Since this process is technically cumbersome, a lighter calibration procedure that could be performed on a more regular basis was put together by the instrument scientists and the instrument engineer responsible. Improving the image quality benefits to everyone, not just high contrast imaging users and therefore this project was given a high priority. The goal here being to provide a good and consistent PSF on night to night basis, not necessarily the best achievable because it would be too time consuming and time conflicting but provide more or less the same PSF: Strehl ratio (SR)$\geqslant 90$\% on the internal PSF fiber (with our reference setup), every night.

\subsection{NAOS calibration and monitoring plan} 
\label{sec:naos}

As part of the \naco ``freezing project'', we initiated a very detailed monitoring and diagnostic plan for the AO arm of \naco (NAOS). Most of the templates and procedures for this existed but it had not always been done in a systematic and regular way, leading to instabilities, inconsistent reliability and frustrating degradation of \naco's original performance level.

\begin{figure}[h!]
\bc
\includegraphics[width=0.70\textwidth]{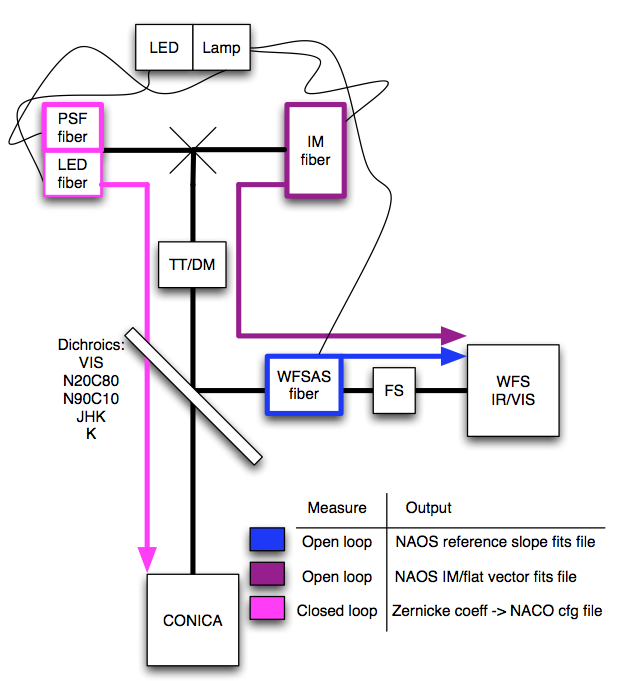}
\caption{Block diagram of \naco. Main elements of \naco are represented, along with the different calibration paths (colored arrows). The blue arrow shows the optical path of NAOS, whose open loop WFS data is saved as the reference slopes. The purple arrow is the interaction matrix (IM) fiber calibration path, it is used for many AO calibrations (including the computation of IM and flat vector files). The pink arrow represents the calibration path for closed-loop CONICA NCPAs measurements.
\label{fig:naos} 
}
\ec
\end{figure} 
 
Figure \ref{fig:naos} summarizes schematically the way \naco's NCPAs can be calibrated thoroughly.
 
Prior to any calibration of the instrument we make sure all four NAOS calibration sources (IM lamp, PSF lamp, WFSAS (WFS artificial source) lamp and LGS IM LED) are stable and within proper flux ranges thanks to our routine monitoring scripts and automatic health check plots (partially implemented).

The following (previously existing) observing blocks (OB) are used to monitor NAOS calibration sources:
\begin{itemize}
\item {\tt WfsTipTilt}: confirm for each WFS in NAOS that the field selector (FS) ``home position'' is well-calibrated in order to acquire the IM fiber. It is equivalent to the determination of the Tip/Tilt NAOS NCPA.
\item {\tt AlignPSF}: confirm for each WFS in NAOS that the FS home position is well-calibrated in order to acquire the PSF fiber. The PSF and IM fibers should appear to the WFS to be at the same position in the image plane.
\item {\tt FocusPSF} and {\tt FocusOffsetPSF}: confirm that PSF and IM fibers are conjugated and that the distance between LED and PSF fibers along the optical axis is correctly calibrated in NAOS.
\item {\tt CheckFocus}: calibrate the defocus aberration of the NAOS NCPA for each dichroic.
\end{itemize}
 
With stable calibration sources, it is possible to perform the procedure described hereafter (and on figure \ref{fig:naos}), which corresponds to the main NAOS NCPA calibration. It is planned to be carried out after any intervention and weekly to monthly depending on the monitoring results.

\begin{itemize}
\item NAOS reference slopes (RS): using the WFSAS, measure the high-order NAOS NCPAs for any given WFS geometry (i.e VIS $14\times14$, $7\times7$ and IR $14\times14$, $7\times7$)
\item Interaction matrices (IM): using the IM fiber, measure the IM for any given WFS geometry. IMs are the conversion matrices from voltages to slopes and are measured efficiently by ``poking'' the deformable mirror (DM) actuators following a Hadamard scheme\cite{meimon2010}.
\end{itemize}
 
 \subsection{CONICA NCPA compensation} 
\label{sec:conica}

Once we are confident that NAOS is well aligned/calibrated, CONICA NCPA (the pink arm on figure \ref{fig:naos}) can be calculated thanks to a phase diversity (PD) algorithm that uses a modulation in the focus term. It used to be done with the so-called {\sl Zernike tool}, a component in the first focal plane wheel of CONICA with pinholes at 0, 1, 2 and 4 mm distance from the true focal plane, which results in re-imaged PSF with appropriate defocus gradient for phase diversity. The whole procedure is very well described in Blanc {\sl et al.}\cite{blanc2003} and Hartung {\sl et al.}\cite{hartung2003pd} (both 2003). In 2010, we noticed the Zernike values introduced in our configuration files were obselete. Indeed, the image quality was neither worst nor better by setting all the values above $Z_{4}$ (focus)  to zero. 

The initial procedure (with the {\sl Zernike tool}) was found to be very time consuming and we experimented with introducing the necessary defocus for PD using the DM directly to introduce a ``pure'' defocus. We first compared the new PD software $\sl OPRA$ ({\sl OTF-based Phase Retrieval Algorithm})  against the original {\sl IDL}-based PD code on the same, 2003 dataset. Results were sensibly identical. Then we achieved similar SRs with one single PD iteration and concluded that our method with the DM was sufficiently good for our SR ranges and needs.

\begin{figure}[h!]
\bc
\includegraphics[width=0.70\textwidth]{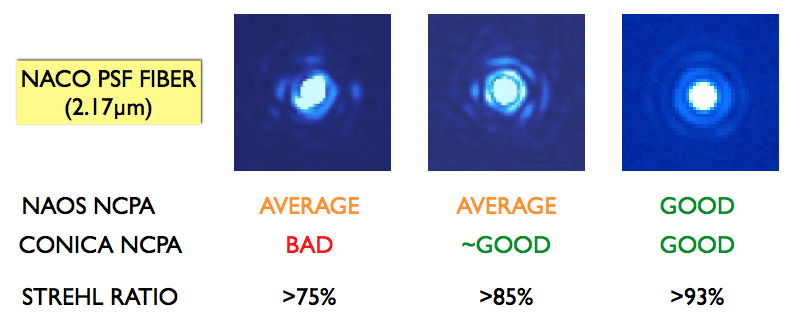}
\caption{Internal image quality in close-loop using the NACO PSF fiber and the narrowband 2.17$\mu m$ filter (our reference).  On the left, one can see the PSF as it was most of the time in 2009-2010 with outdated CONICA NCPA and probably not so good NAOS calibrations. In the centre, the $\sl OPRA$  corrected PSF (probably correcting both NAOS and CONICA NCPAs at once). On the right, the optimized PSF in November 2011 after performing all calibrations (NAOS and CONICA separately).
\label{fig:psf} 
}
\ec
\end{figure}

By doing many setup combinations, one would determine the Zernike values for each CONICA  element (objective, filter, etc.) and NAOS dichroic. We decided to assume that most NCPA came from the main collimator (common to all setups), from the objective and that differential NCPAs from one dichroic to another were negligible (thus avoiding the setup of an external source to pass by the either dichroic). In this way the generic setup (VIS/S13/$Br\gamma$) values were applied to all S13 filters and only checked/changed $Z_{4}$ (focus) for other filters. This was decided after performing many tests and it was found to be adequate for our needs. $Br\gamma$ results are shown on figure \ref{fig:psf}.

Finally, we noticed (by running {\sl OPRA}) regularly when NAOS was appropriately calibrated that CONICA NCPA were rather stable with time as expected since CONICA is cold. On the contrary NAOS NCPA are very much subject to temperature gradients (from day to night and from seasonal transitions). Rotator angles have some effect on the NCPA but it is not dramatic and therefore not necessary to implement a special scheme.

\begin{figure}[h!]
\bc
\includegraphics[width=0.90\textwidth]{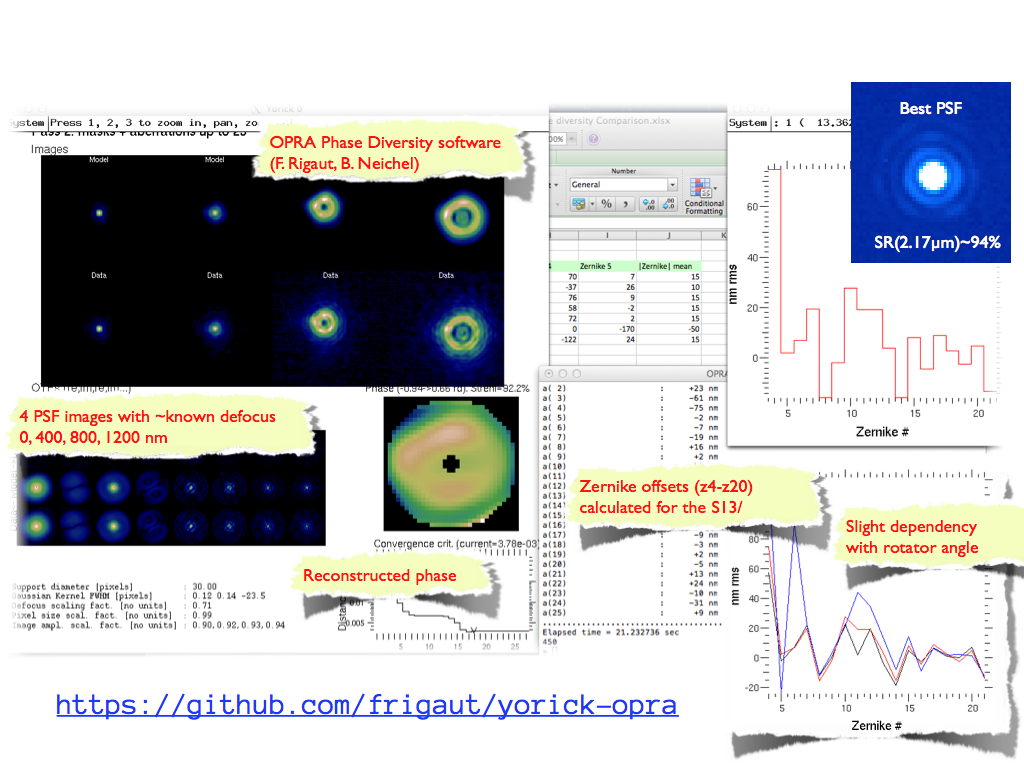}
\caption{Annotated screenshot of the $\sl OPRA$  phase diversity software ran on \naco data. On the top left panel one can see the four images (bottom) taken with 0, 400, 800, 1200 nm of defocus applied on the DM and the associated four fitted models (top). Slightly lower are the residual and the reconstructed wavefront variance map (the bigger picture in the middle). On the top right corner is a plot of the Zernike offsets (amplitude in nm versus order) to be applied to the DM to correct the wavefront and hence the PSF with the best PSF achieved with a SR$\sim$94$\pm$3\% after 2 iterations. The plot on the bottom right corner shows various curves of the Zernike offsets estimated at 0, 45 and 90$\deg$ Rotator angles. Any offset $\leqslant$20 nm was judged to have a negligible influence on the results and therefore we only considered offsets up to $Z_{12}$ on Noll's basis\cite{noll76}. The current $\sl github \; URL$ of the $\sl OPRA$  package is given.
\label{fig:opra} 
}
\ec
\end{figure}

Here is the summary of the now called ``{\sl OPRA}'' procedure. A screen shot of the PD steps is described on figure \ref{fig:opra}:

\bl
\item CONICA NCPA: using the {\tt cnstooTakePDData} script and {\sl OPRA}, determine the set of high- order aberration coefficients ($Z_{i}\geqslant 5$) common to CONICA setups, with the narrowband (NB) filters, with the VIS $14\times14$ NAOS mode (our reference setups) and the S13 (13 \mas/pixel) objective (\naco's finest camera). 
\item Configuration files update.
\item Focus optimization: check and adjust the defocus term ($Z_{4}$) for each filter manually. Run {\sl OPRA} if necessary. 
\item Check and eventual iteration(s) to make sure the most common combinations reach the SR thresholds.
\el

At this stage, the system is now calibrated for the visible WFS modes, and is roughly calibrated for the others. The next logical step is then the extension of this procedure to the IR WFS (different dichroics) and the LGS modes which requires similar actions.

All of the above is only valid for the wavelength range for which we have calibration lamps working: 1-2.5 $\mu$m ({\sl aka} SW for short wavelengths). For long wavelengths ({\sl aka} LW) and the two corresponding objectives (L27 and L54), it is necessary to perform the PSD/{\sl OPRA} step on-sky using a bright star. This project is on-going and has lower priority since the SR regime we are talking about  for the most used band (L', centered at 3.8$\mu$m) is already 70 to 85 \% even with approximative NAOS/CONICA NCPA calibrations. We might require a different PD treatment than {\sl OPRA}'s to overcome the remaining aberrations for the LW.

\subsection{First on-sky phase retrieval demonstration using NACO}

Riaud {\sl et al.}(2012, accepted)\cite{riaud2012} used an original implementation of the Nijboer-Zernike phase retrieval method to measure the phase and amplitude of static wavefront errors generated by non-common path aberrations between the NAOS dichroic and the CONICA detector. The originality of the demonstration lies into the online on-sky nature of the retrieval. Indeed, such calibrations are usually done during the daytime on internal calibration sources. Here, the authors used three images (intra, in and extra focus) of a real star - HD25026 - acquired in the Br$\gamma$ filter (2.17$\mu m$) in closed loop and temporally averaged over 30 seconds to derive the complex amplitude of the \naco pupil in quasi real-time. The measurement yielded 0.105 wave rms (equivalent to 60\% Strehl ratio) for the phase, and about 10\% of amplitude variations, consistent with other similar measurements. The result of this ground-breaking experiment demonstrated the usefulness of online phase retrieval, which is relevant in the context of next-generation high contrast imagers such as SPHERE. Online on-sky quasi-real time measurements will indeed be needed to mitigate the time variability of quasi-static speckles (due to temperature and gravity vector changes) and to increase observing efficiency, both leading to better image quality.

\section{CURRENT AO PERFORMANCES} 
\label{sec:perf}

It is always difficult to talk about nominal on-sky AO performances because they depend highly on the atmospheric conditions, airmass, guide star distance and magnitude, sky transparency and WFS frame rate (or AO close-loop bandwidth associated to the guide star magnitude). In the case of diffraction limited imaging, the Strehl ratio (SR) is a good metric (because significantly bigger than a couple percents). In the case of very high contrasts, the best metric becomes the intrinsic contrast at a given separation of the residual wavefront error measured by the WFS in nm.
\naco's performances are often misinterpreted because statistics are usually performed using automatic Strehl ratios (SR) measurements on photometric standards, most of which are acquired using low WFS frame rate (60 or 120 Hz in the visible) on rather faint stars. In this paper, we prefer to talk about ``peak performances'' with good but not exceptional conditions to gauge whether the system is performing at least as good than when commissioned by AO experts.
  
\subsection{On-sky AO performances} 
\label{sec:skyperf}

Recent top performances (with seeing $\leqslant0.6$\arcsec and $tau_{0}\geqslant5$ ms) match the performances that were obtained at commissioning\cite{rousset2003}:
\begin{itemize}
\item SR$\sim60\pm5$\% Ks-band (2.18$\mu$ m) with NGS with  IR-WFS\cite{gendron2003} ($14\times 14$,162 Hz)
\item SR$\sim60\pm5$\% Ks-band with NGS with VIS-WFS ($14\times 14$, 444 Hz) and $\sim 40\pm4$\% in H-band 
\item SR$\sim80\pm5$\% L'-band  with NGS with either WFS
\item SR$\sim35\pm4$\% Ks-band with LGS ($14\times 14$,  120 or 240 Hz).
\end{itemize}

To estimate our SRs we used several ''Strehl Meters'' and tested them against each other as was done in the {\sl Is That Really Your Strehl Ratio?} 2004 paper \cite{roberts2004}. These are what we call ``raw'' SRs because they correspond to a few second long exposure. With our fast cube mode (all short exposure frames are saved) one can recenter, select and stack frames an gain easily 10\% on the SR. 

The {\sl Abism (Adaptative Background Interactive Strehl Meter)} {\sl Python}-based software was then developed to match \naco's operational needs and was extensively tested to be used for all measurements. It is in the process of being fully integrated to the operations scheme to grade AO observations based on AO performances. This software should provide user independent results (Strehl ratio, full width at half maximum, encircled energy) thanks to innovative ways of determining the background (therefore the photometry) and fitting the somewhat isolated PSF reference star with an adaptive 2-D function (Gaussian, Moffat or Bessel, depending on the level of AO correction).

\subsection{Limitations for higher contrasts} 
 \label{sec:lim}
 
\begin{figure}[h!]
\bc
\includegraphics[width=0.85\textwidth]{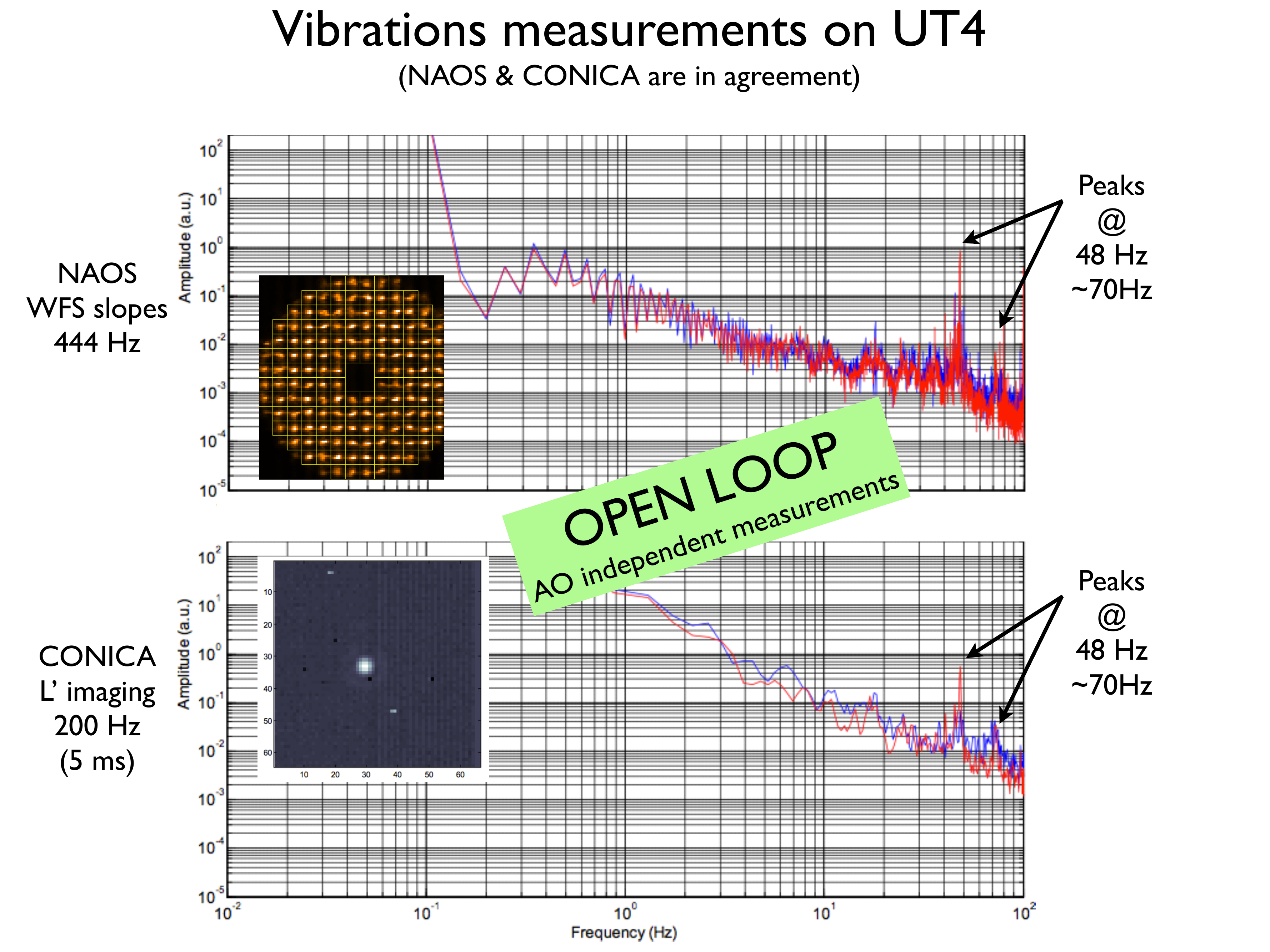}
\caption{UT4 vibrations as seen by NAOS (SH WFS slopes measured at 444Hz) and CONICA (centro\"{i}ds measured on a bright star at 200 Hz) in open-loop to avoid any damping from the AO and at 4$\mu$m to have an almost diffraction limited image.
\label{fig:vib} 
}
\ec
\end{figure}

The correction of quasi-static NCPA addresses the PSF shape and intrinsic Strehl ratio. However, there are many other limitations to optimal image quality, and high contrasts. Our practical experience with NACO yields the following non-exhaustive list:
\begin{itemize}
\item AO resolution elements: \naco with only $14\times14$ sub-apertures and an 185-element deformable mirror (DM) cannot perform much better than its nominal yield.
\item Coherence time: large high contrast imaging surveys on tens of stars have proven that one of the most influential parameters is $\tau_{0}$, the atmospheric coherence time. Indeed, \naco visible SH WHS runs at 480 Hz at most and often under-samples the atmospheric signal (when there are high winds, {\sl jet-stream}. Contrast curves acquired with $\tau_{0}$ varying from 1 ms to 10 ms show that when $tau_{0}\geqslant$5 ms, post-processing ADI techniques (i.e LOCI\cite{lafreniere2007}) do not improve dramatically the inner-working-angle (IWA), which is already good with such conditions and a large ($\geqslant 40 \deg$) amount of field rotation. On the other hand, when $\tau_{0}\leqslant$2 ms, these techniques can help efficiently to recover the best possible IWA.
\item Telescope vibrations: already spotted in the early days of \naco, especially a $\sim$48 Hz, we characterized the UT4 vibrations using both CONICA centro\"{i}ding at 200 Hz and NAOS slopes at 444 Hz in open-loop as shown on figure \ref{fig:vib}.  

\item Telescope spiders, central obstruction and diffraction aigrets: addressed by various groups\cite{serabyn2007}, off-axis telescope may very well be the future of high contrast techniques and coronagraphs. 
\item Ghosts handling: this is definitely a problem for long slit spectroscopy using \naco\cite{janson2010}. High contrast spectrographs need special care for reducing ghost from the design and integration stages.
\end{itemize}

\section{Pupil tracking drift: solved!} 
\label{sec:drift}

Pupil tracking (PT) was implemented originally (in 2007) for the SAM mode of \naco which makes use of pupil plane masks with holes. These holes act as independent telescope apertures, only the light is combined in a ``Fizeau'' way, and a 2-D interferometric pattern is seen in the image plane on CONICA's detector. At a fast frame rate, visibilities for each non-redundant baseline can be measured as well as closure phases on each triplet. The first implementation of the pupil tracking had only one purpose, that was to align the pupil spiders and central obscuration with the SAM mask so that its holes never cross the shadows of these telescopes support structures.

The implementation and routine use of angular differential imaging\cite{marois2006} on \naco came as a byproduct of SAM and PT as soon as this technique proved to be superior to PSF subtraction for low mass companion searches and characterization.

Since the PT implementation had focused on the pupil alignment, not all the software was PT ``compliant'' and the guide star PSF - supposedly the center of rotation - was describing a circle at the speed of the parallactic angle variation (azimuth axis of the telescope). After investigating for nearly two years, we finally noticed that the field selector (FS: two mirrors that move along to keep the star in the WFS no matter how the telescope tracks and offsets) was incorrectly driven in the case of PT. In normal field tracking (FT), it has to update its position with respect to the azimuth angle. In PT, this offload should have been turned off. Since every offset introduced on the FS is seen by the SH WFS and drives the tip/tilt mirror (TTM), we had this circular drift issue that prevented us to efficiently use any focal plane masks (i.e 4QPM and opaque Lyot coronagraphs).

Thanks to cube mode and saturated imaging (sort of ``electronic coronagraph''), people still managed to do high contrast imaging efficiently with \naco, especially in the L'-band where the SRs are high and the background limited performances of \naco are perhaps better than its competitors\cite{quanz2012}.

\begin{figure}[h!]
\bc
\includegraphics[width=0.45\textwidth]{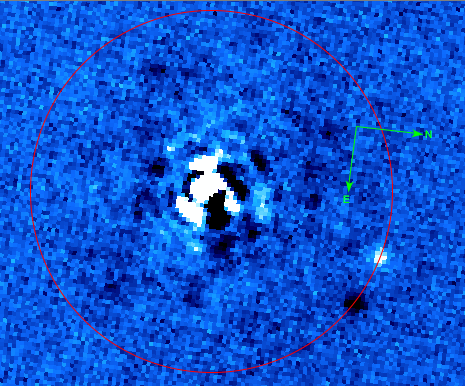}
\caption{Since the infamous PT drift has been solved, the observer witnesses ``real time ADI'' as the first frame can be subtracted to the current frame and one can see relatively faint companions (or field star) rotating. Here is an example with a 10 to 1 contrast binary star and about 15 degrees of field rotation. The black features correspond  to the negative, subtracted frame.
\label{fig:drift} 
}
\ec
\end{figure}

Since October 15th 2011, the drift is fixed as shown on figure \ref{fig:drift}, focal plane coronagraphs (Lyot, 4QPM and possibly the AGPM) are pertinent and fully compatible with PT! The residual drift of $\sim$20 \mas/hour correspond to an imperfect compensation of the mechanical flexures at different angles. It is totally manageable in practice (e.g.~manual recentering every 10-20 minutes depending on the coronagraph). 

We also checked that atmospheric differential refraction compensation was working using a large wavelength baseline (i.e L' filter on CONICA and the VIS WFS) at an airmass $\geqslant$ 1.8. No additional drift or image elongation was sensed after making sure the NAOS software always uses the current observing wavelength.

\section{Improvements to SAM interferometry}
\label{sec:sam}

The SAM mode is very interesting for small angle search for companions\cite{huelamo2011} in the $\geqslant$0.3 to 5 $\lambda/D$ region thanks to model fitting\cite{lacour2011_2}, field rotation and as for ADI, precise astrometry versus proper motion to confirm the bound nature of the putative candidate low mass companion (which can be a background star or galaxy or a bright feature in a disk for example).

In 2011 a strategy of ``star hopping'' was put together\cite{lacour2011} in order to switch faster from the science target to one or several PSF calibrators. The VLT being an active telescope (with active optics), many time variant parameters can influence the optical transfer function (OTF). In addition, the atmospheric transfer function is evolving rapidly as well (what counts here is the residual, post-AO wavefront variance on the full pupil that translates mainly into differential piston between the sparse apertures) and therefore it was very difficult to calibrate the PSF with sufficient time resolution. With star ''hopping'', one can ``hop'' from the science target to a calibrator much faster and within the same observing block thanks to carefully calculated offsets. Provided that the calibrator is well chosen (within a degree and of the same brightness than the science target), one can close the AO loop on it without reacquiring with the WFS, making the process efficient and stable. This strategy has proven to improve drastically the results on SAM ever since in terms of detection limits (for example, on a bright star, a  7.5 magnitude uniform contrast can be reached at 50-500 \mas in L'-band).

\section{Perspectives} 
\label{sec:ideas}

In October/November 2012, a 6-week maintenance intervention will allow us to improve \naco in several ways and maintain it at its best level during the last few semesters of operations. An upgrade of the laser guide star facility (LGSF) should occur also in the next months to make the system more robust and to take advantage of the new, larger field of view SH WFS\cite{girard2010spie, kasper2010lgs}.

We hope to resume operations quickly and successfully and keep \naco healthy for a few more periods to eventually benefit from its 3-5$\mu$m complementary capability when SPHERE will be running. 

Among the proposals is to introduce one last coronagraph in CONICA, one that is particularly appealing. It is a newly manufactured AGPM, optimized at 4.05 $\mu$m. It is the first of its kind. The VLT (UT3/VISIR at 10$\mu$m and
UT4/\naco at 4$\mu$m) would have the first 8-meter class telescopes to be equipped with such high
performance coronagraph which would benefit fully from the absence of pupil tracking drift (section \ref{sec:drift}), the NCPA compensated PSF (section \ref{sec:iq}) and would open a new parameter space, complementary to that of SPHERE, GPI and most of the new generation of ``planet imagers''. This proposal is under review but simulations are very encouraging.

Since October 2011 we also have a new {\sl GRAB} button which enables us to save RTC telemetry data (up to 4096 consecutive slopes slopes or $\sim$9 seconds of data at 444 Hz) simultaneously with high frame - cube mode - CONICA images. Having both pupil and image plane information synchronized allows to perform phase retrieval and focal plane wavefront sensing techniques such as the {\sl Phase Sorting Interferometry}\cite{otten2012spie}. Preliminary tests on the APP PSF are encouraging. It is also very useful to take vibrations measurements.

\section{Conclusion} 
\label{sec:life}

It is difficult to conceive instruments, difficult to design and tolerance instruments, difficult to test and integrate instruments, commission  and put them into operation. It is also difficult to maintain top performance of complex instruments during years of operations with constant staff turnover and limited resources. \naco is a state-of-the-art AO  fed, world class instrument with over 10 years of successful operations. We are still attempting to improve its efficiency and, as {\sl Paranal AO Operations Group}, to learn with it towards the future AO instruments and facilities.

\acknowledgments     %>>>> equivalent to \section*{ACKNOWLEDGMENTS}       
 
Julien Girard would like to acknowledge the whole \naco IOT in particular Lowell Tacconi-Garman who provided very useful feedback on the elaboration of this paper. Finally he thanks ESO Paranal management, Engineering  and IOT Coordinator Alain Smette for their valuable support. We are also thankful to Markus Hartung who shared his experience with us, Fran\c cois Rigaut and Damien Gratadour who worked on 
%{\tt \href{https://github.com/frigaut/yorick-opra}{
$\sl Yorick/OPRA$
with Benoit Neichel.

\bibliographystyle{spiebib}   %>>>> makes bibtex use spiebib.bst
\bibliography{biblio_spie2012}

\begin{thebibliography}{10}

\bibitem{beuzit2010lyot}
{Beuzit}, J.-L., {Feldt}, M., {Mouillet}, D., {Dohlen}, K., {Puget}, P.,
  {Wildi}, F., and {SPHERE Consortium}, ``{SPHERE: a planet imager for the
  VLT},'' in [{\em In the Spirit of Lyot 2010}{\nolinebreak\hspace{0.1em}]},
  (Oct. 2010).

\bibitem{kerber2010spie}
{Kerber}, F., {Kaeufl}, H.~U., {van den Ancker}, M., {Baksai}, P., {Dubreuil},
  D., {Durand}, G., {Dobrzycka}, D., {Finger}, G., {Hummel}, C., {Ives}, D.,
  {Jakob}, G., {Lagadec}, E., {Lundin}, L., {Marconi}, G., {Moerchen}, M.,
  {Momany}, Y., {Nuernberger}, D., {Pantin}, E., {Riquelme}, M.,
  {Siebenmorgen}, R., {Smette}, A., {Venema}, L., {Weilenmann}, U., and
  {Yegorova}, I., ``{Upgrade of VISIR the mid-infrared instrument at the
  VLT},'' in [{\em Society of Photo-Optical Instrumentation Engineers (SPIE)
  Conference Series}{\nolinebreak\hspace{0.1em}]},  {\em Society of
  Photo-Optical Instrumentation Engineers (SPIE) Conference Series} {\bf 7735}
  (July 2010).

\bibitem{marchetti2012spie}
{Marchetti}, E., {Le Louarn}, M., {Fedrigo}, E., {Soenke}, C., {Madec}, P., and
  {Hubin}, N., ``{ERIS adaptive optics system design},'' in [{\em {Adaptive
  Optics Systems III, Proceedings of the SPIE}}{\nolinebreak\hspace{0.1em}]},
  {Ellerbroek}, B., {Marchetti}, E., and {V\'{e}ran}, J.-P., eds.,  {\bf -}
  (July 2012).

\bibitem{lenzen2003}
{Lenzen}, R., {Hartung}, M., {Brandner}, W., {Finger}, G., {Hubin}, N.~N.,
  {Lacombe}, F., {Lagrange}, A., {Lehnert}, M.~D., {Moorwood}, A.~F.~M., and
  {Mouillet}, D., ``{NAOS-CONICA first on sky results in a variety of observing
  modes},'' in [{\em Society of Photo-Optical Instrumentation Engineers (SPIE)
  Conference Series}{\nolinebreak\hspace{0.1em}]},  {M.~Iye \&
  A.~F.~M.~Moorwood}, ed., {\em Society of Photo-Optical Instrumentation
  Engineers (SPIE) Conference Series} {\bf 4841},  944--952 (Mar. 2003).

\bibitem{rousset2003}
{Rousset}, G., {Lacombe}, F., {Puget}, P., {Hubin}, N.~N., {Gendron}, E.,
  {Fusco}, T., {Arsenault}, R., {Charton}, J., {Feautrier}, P., {Gigan}, P.,
  {Kern}, P.~Y., {Lagrange}, A., {Madec}, P., {Mouillet}, D., {Rabaud}, D.,
  {Rabou}, P., {Stadler}, E., and {Zins}, G., ``{NAOS, the first AO system of
  the VLT: on-sky performance},'' in [{\em Society of Photo-Optical
  Instrumentation Engineers (SPIE) Conference
  Series}{\nolinebreak\hspace{0.1em}]},  {P.~L.~Wizinowich \&amp;
  D.~Bonaccini}, ed., {\em Society of Photo-Optical Instrumentation Engineers
  (SPIE) Conference Series} {\bf 4839},  140--149 (Feb. 2003).

\bibitem{brandner2002}
{Brandner}, W., {Rousset}, G., {Lenzen}, R., {Hubin}, N., {Lacombe}, F.,
  {Hofmann}, R., {Moorwood}, A., {Lagrange}, A., {Gendron}, E., {Hartung}, M.,
  {Puget}, P., {Ageorges}, N., {Biereichel}, P., {Bouy}, H., {Charton}, J.,
  {Dumont}, G., {Fusco}, T., {Jung}, Y., {Lehnert}, M., {Lizon}, J., {Monnet},
  G., {Mouillet}, D., {Moutou}, C., {Rabaud}, D., {R{\"o}hrle}, C., {Skole},
  S., {Spyromilio}, J., {Storz}, C., {Tacconi-Garman}, L., and {Zins}, G.,
  ``{NAOS+CONICA at YEPUN: first VLT adaptive optics system sees first
  light},'' {\em The Messenger}~{\bf 107},  1--6 (Mar. 2002).

\bibitem{girard2010spie}
{Girard}, J.~H.~V., {Kasper}, M., {Quanz}, S.~P., {Kenworthy}, M.,
  {Rengaswamy}, S., {Schoedel}, R., {Dobrzycka}, D., {Gallenne}, A.,
  {Gillessen}, S., G., G., {Huerta}, N., {Kervella}, P., {Kornweibel}, N.,
  {Lenzen}, R., {Lundin}, L., {M\'{e}rand}, A., {Montagnier}, G., {O'Neal}, J.,
  {Witzel}, G., and {Zins}, G., ``{Status and new operation modes of the
  versatile VLT/NaCo},'' in [{\em {Adaptive Optics Systems II, Proceedings of
  the SPIE}}{\nolinebreak\hspace{0.1em}]},  {Hubin}, N., E., M.~C., and
  {Wizinowich}, P.~L., eds., {\em Astronomical Instrumentation} {\bf 7736},
  Society of Photo-Optical Instrumentation Engineers (SPIE) Conference Series
  (July 2010).

\bibitem{schoedel2002}
{Sch\"{o}del}, R., {Ott}, T., {Genzel}, R., {Hofmann}, R., {Lehnert}, M.,
  {Eckart}, A., {Mouawad}, N., {Alexander}, T., {Reid}, M.~J., {Lenzen}, R.,
  {Hartung}, M., {Lacombe}, F., {Rouan}, D., {Gendron}, E., {Rousset}, G.,
  {Lagrange}, A., {Brandner}, W., {Ageorges}, N., {Lidman}, C., {Moorwood},
  A.~F.~M., {Spyromilio}, J., {Hubin}, N., and {Menten}, K.~M., ``{A star in a
  15.2-year orbit around the supermassive black hole at the centre of the Milky
  Way},'' {\em \nat}~{\bf 419},  694--696 (Oct. 2002).

\bibitem{gillessen2009}
{Gillessen}, S., {Eisenhauer}, F., {Trippe}, S., {Alexander}, T., {Genzel}, R.,
  {Martins}, F., and {Ott}, T., ``{Monitoring Stellar Orbits Around the Massive
  Black Hole in the Galactic Center},'' {\em \apj}~{\bf 692},  1075--1109 (Feb.
  2009).

\bibitem{chauvin2004}
{Chauvin}, G., {Lagrange}, A., {Dumas}, C., {Zuckerman}, B., {Mouillet}, D.,
  {Song}, I., {Beuzit}, J., and {Lowrance}, P., ``{A giant planet candidate
  near a young brown dwarf. Direct VLT/NACO observations using IR wavefront
  sensing},'' {\em \aap}~{\bf 425},  L29--L32 (Oct. 2004).

\bibitem{eisenhauer2005}
{Eisenhauer}, F., {Genzel}, R., {Alexander}, T., {Abuter}, R., {Paumard}, T.,
  {Ott}, T., {Gilbert}, A., {Gillessen}, S., {Horrobin}, M., {Trippe}, S.,
  {Bonnet}, H., {Dumas}, C., {Hubin}, N., {Kaufer}, A., {Kissler-Patig}, M.,
  {Monnet}, G., {Str{\"o}bele}, S., {Szeifert}, T., {Eckart}, A.,
  {Sch{\"o}del}, R., and {Zucker}, S., ``{SINFONI in the Galactic Center: Young
  Stars and Infrared Flares in the Central Light-Month},'' {\em \apj}~{\bf
  628},  246--259 (July 2005).

\bibitem{eckart2008}
{Eckart}, A., {Baganoff}, F.~K., {Zamaninasab}, M., {Morris}, M.~R.,
  {Sch{\"o}del}, R., {Meyer}, L., {Muzic}, K., {Bautz}, M.~W., {Brandt}, W.~N.,
  {Garmire}, G.~P., {Ricker}, G.~R., {Kunneriath}, D., {Straubmeier}, C.,
  {Duschl}, W., {Dovciak}, M., {Karas}, V., {Markoff}, S., {Najarro}, F.,
  {Mauerhan}, J., {Moultaka}, J., and {Zensus}, A., ``{Polarized NIR and X-ray
  flares from Sagittarius A*},'' {\em \aap}~{\bf 479},  625--639 (Mar. 2008).

\bibitem{janson2010}
{Janson}, M., {Bergfors}, C., {Goto}, M., {Brandner}, W., and {Lafreni{\`e}re},
  D., ``{Spatially Resolved Spectroscopy of the Exoplanet HR 8799 c},'' {\em
  \apjl}~{\bf 710},  L35--L38 (Feb. 2010).

\bibitem{mawet2012spie}
{Mawet}, D., {Pueyo}, L., {Lawson}, P., {Mugnier}, L., {Traub}, W.,
  {Boccaletti}, A., {Trauger}, J., {Gladysz}, S., {Serabyn}, E., {Milli}, J.,
  {Belikov}, R., {Kasper}, M., {Baudoz}, P., {Macintosh}, B., {Marois}, C.,
  {Oppenheimer}, B., {Barrett}, H., {Beuzit}, J.-L., {Devaney}, N., {Girard},
  J., {Guyon}, O., {Krist}, J., {Mennesson}, B., {Mouillet}, D., {Murakami},
  N., {Poyneer}, L., {Savransky}, D., {V\'{e}rinaud}, C.~V., and {Wallace},
  J.~K., ``{Review of small-angle coronagraphic techniques in the wake of
  ground-based second-generation adaptive optics systems},'' in [{\em
  {Ground-based and Airborne Instrumentation for Astronomy IV, Proceedings of
  the SPIE}}{\nolinebreak\hspace{0.1em}]},  {McLean}, I., {Ramsay}, S., and
  {Takami}, H., eds.,  {\bf -} (July 2012).

\bibitem{kasper2012spie}
{Kasper}, M., ``{Adaptive optics for high contrast imaging},'' in [{\em
  {Adaptive Optics Systems III, Proceedings of the
  SPIE}}{\nolinebreak\hspace{0.1em}]},  {Ellerbroek}, B., {Marchetti}, E., and
  {V\'{e}ran}, J.-P., eds.,  {\bf -} (July 2012).

\bibitem{telbib}
{ESO Library}, {\em {ESO Telescope Bibliography}}.
\newblock European Southern Observatory (http://telbib.eso.org) (2012).

\bibitem{paufique2010spie}
{Paufique}, J., {Bruton}, A., {Glindemann}, A., {Jost}, A., {Kolb}, J.,
  {Jochum}, L., {Le Louarn}, M., {Kiekebusch}, M., {Hubin}, N., {Madec}, P.-Y.,
  {Conzelmann}, R., {Siebenmorgen}, R., {Donaldson}, R., {Arsenault}, R., and
  {Tordo}, S., ``{GRAAL: a seeing enhancer for the NIR wide-field imager
  Hawk-I},'' in [{\em Society of Photo-Optical Instrumentation Engineers (SPIE)
  Conference Series}{\nolinebreak\hspace{0.1em}]},  {\em Society of
  Photo-Optical Instrumentation Engineers (SPIE) Conference Series} {\bf 7736}
  (July 2010).

\bibitem{stuik2006}
{Stuik}, R., {Bacon}, R., {Conzelmann}, R., {Delabre}, B., {Fedrigo}, E.,
  {Hubin}, N., {Le Louarn}, M., and {Str\"{o}bele}, S., ``{GALACSI The ground
  layer adaptive optics system for MUSE},'' {\em \nar}~{\bf 49},  618--624
  (Jan. 2006).

\bibitem{sauvage2010spie}
{Sauvage}, J.-F., {Fusco}, T., {Petit}, C., {Meimon}, S., {Fedrigo}, E.,
  {Suarez Valles}, M., {Kasper}, M., {Hubin}, N., {Beuzit}, J.-L., {Charton},
  J., {Costille}, A., {Rabou}, ., P., {Mouillet}, D., {Baudoz}, P., {Buey}, T.,
  {Sevin}, A., {Wildi}, F., and {Dohlen}, K., ``{SAXO, the eXtreme Adaptive
  Optics System of SPHERE: overview and calibration procedure},'' in [{\em
  Society of Photo-Optical Instrumentation Engineers (SPIE) Conference
  Series}{\nolinebreak\hspace{0.1em}]},  {\em Society of Photo-Optical
  Instrumentation Engineers (SPIE) Conference Series} {\bf 7736} (July 2010).

\bibitem{fedrigo2010spie}
{Fedrigo}, E., {Bourtembourg}, R., {Donaldson}, R., {Soenke}, C., {Suarez
  Valles}, M., and {Zampieri}, S., ``{SPARTA for the VLT: status and plans},''
  in [{\em Society of Photo-Optical Instrumentation Engineers (SPIE) Conference
  Series}{\nolinebreak\hspace{0.1em}]},  {\em Society of Photo-Optical
  Instrumentation Engineers (SPIE) Conference Series} {\bf 7736} (July 2010).

\bibitem{suarezvalles2012spie}
{Su\'{a}rez Valles}, M., {Fedrigo}, E., {Donaldson}, R., {Soenke}, C.,
  {Zampieri}, S., {Bourtembourg}, R., and {Tischer}, T., ``{SPARTA for the VLT:
  status and plans},'' in [{\em {Adaptive Optics Systems III, Proceedings of
  the SPIE}}{\nolinebreak\hspace{0.1em}]},  {Ellerbroek}, B., {Marchetti}, E.,
  and {V\'{e}ran}, J.-P., eds.,  {\bf -} (July 2012).

\bibitem{kasper2005}
{Kasper}, M., {Ageorges}, N., {Boccaletti}, A., {Brandner}, W., {Close}, L.~M.,
  {Davies}, R., {Finger}, G., {Genzel}, R., {Hartung}, M., {Kaufer}, A.,
  {Kellner}, S., {Hubin}, N., {Lenzen}, R., {Ludman}, C., {Monnet}, G.,
  {Moorwood}, A., {Ott}, T., {Riaud}, P., {Roser}, H., {Rouan}, D., and
  {Spyromilio}, J., ``{New observing modes of NACO},'' {\em The Messenger}~{\bf
  119},  11--+ (Mar. 2005).

\bibitem{biller2004spie}
{Biller}, B.~A., {Close}, L., {Lenzen}, R., {Brandner}, W., {McCarthy}, D.~W.,
  {Nielsen}, E., and {Hartung}, M., ``{Suppressing speckle noise for
  simultaneous differential extrasolar planet imaging (SDI) at the VLT and
  MMT},'' in [{\em Society of Photo-Optical Instrumentation Engineers (SPIE)
  Conference Series}{\nolinebreak\hspace{0.1em}]},  {Bonaccini Calia}, D.,
  {Ellerbroek}, B.~L., and {Ragazzoni}, R., eds., {\em Society of Photo-Optical
  Instrumentation Engineers (SPIE) Conference Series} {\bf 5490},  389--397
  (Oct. 2004).

\bibitem{boccaletti2004}
{Boccaletti}, A., {Riaud}, P., {Baudoz}, P., {Baudrand}, J., {Rouan}, D.,
  {Gratadour}, D., {Lacombe}, F., and {Lagrange}, A.-M., ``{The Four-Quadrant
  Phase Mask Coronagraph. IV. First Light at the Very Large Telescope},'' {\em
  \pasp}~{\bf 116},  1061--1071 (Nov. 2004).

\bibitem{tuthill2010spie}
{Tuthill}, P., {Lacour}, S., {Amico}, P., {Ireland}, M., {Norris}, B.,
  {Stewart}, P., {Evans}, T., {Kraus}, A., {Lidman}, C., {Pompei}, E., and
  {Kornweibel}, N., ``{Sparse Aperture Masking (SAM) at NAOS/CONICA on the
  VLT},'' {\em ArXiv e-prints}~{\bf 7735} (June 2010).

\bibitem{kasper2009}
{Kasper}, M., {Amico}, P., {Pompei}, E., {Ageorges}, N., {Apai}, D.,
  {Argomedo}, J., {Kornweibel}, N., and {Lidman}, C., ``{Direct Imaging of
  Exoplanets and Brown Dwarfs with the VLT: NACO Pupil-stabilised Lyot
  Coronagraphy at 4 {$\mu$}m},'' {\em The Messenger}~{\bf 137},  8--13 (Sept.
  2009).

\bibitem{codona2006}
{Codona}, J.~L., {Kenworthy}, M.~A., {Hinz}, P.~M., {Angel}, J.~R.~P., and
  {Woolf}, N.~J., ``{A high-contrast coronagraph for the MMT using phase
  apodization: design and observations at 5 microns and 2 {$\lambda$}/D
  radius},'' in [{\em Society of Photo-Optical Instrumentation Engineers (SPIE)
  Conference Series}{\nolinebreak\hspace{0.1em}]},  {\em Society of
  Photo-Optical Instrumentation Engineers (SPIE) Conference Series} {\bf 6269}
  (July 2006).

\bibitem{quanz2010}
{Quanz}, S.~P., {Meyer}, M.~R., {Kenworthy}, M.~A., {Girard}, J.~H.~V.,
  {Kasper}, M., {Lagrange}, A.-M., {Apai}, D., {Boccaletti}, A., {Bonnefoy},
  M., {Chauvin}, G., {Hinz}, P.~M., and {Lenzen}, R., ``{First Results from
  Very Large Telescope NACO Apodizing Phase Plate: 4 {$\mu$}m Images of The
  Exoplanet {$\beta$} Pictoris b},'' {\em \apjl}~{\bf 722},  L49--L53 (Oct.
  2010).

\bibitem{quanz2011}
{Quanz}, S.~P., {Kenworthy}, M.~A., {Meyer}, M.~R., {Girard}, J.~H.~V., and
  {Kasper}, M., ``{Searching for Gas Giant Planets on Solar System Scales: VLT
  NACO/APP Observations of the Debris Disk Host Stars HD172555 and HD115892},''
  {\em \apjl}~{\bf 736},  L32 (Aug. 2011).

\bibitem{kenworthy2010spie}
{Kenworthy}, M., {Meyer}, M., {Quanz}, S.~P., {Kasper}, M., {Lenzen}, R.,
  {Codona}, J., {Girard}, J. H.~V., and {Hinz}, P., ``{An Apodizing Phase Plate
  Coronagraph for the VLT},'' in [{\em {Ground-based and Airborne
  Instrumentation for Astronomy III, Proceedings of the
  SPIE}}{\nolinebreak\hspace{0.1em}]},  {McLean}, I.~S., K., R.~S., and
  {Takami}, H., eds., {\em Astronomical Instrumentation} {\bf 7735}, Society of
  Photo-Optical Instrumentation Engineers (SPIE) Conference Series (July 2010).

\bibitem{girard2010lyot}
{Girard}, J.~H.~V., {Janson}, M., {Quanz}, S.~P., {Kenworthy}, M.~A., {Meyer},
  M.~R., {Kasper}, M., {Lenzen}, R., and {Wehmeier}, U., ``{Coronagraphic
  Upgrades at the VLT/NaCo: 4-Micron APP Enhanced Spectroscopy?},'' in [{\em In
  the Spirit of Lyot 2010}{\nolinebreak\hspace{0.1em}]},  (Oct. 2010).

\bibitem{lacour2011}
{Lacour}, S., {Tuthill}, P., {Ireland}, M., {Amico}, P., and {Girard}, J.,
  ``{Sparse aperture masking on Paranal},'' {\em The Messenger}~{\bf 146},
  18--23 (Dec. 2011).

\bibitem{mawet2005}
{Mawet}, D., {Riaud}, P., {Absil}, O., and {Surdej}, J., ``{Annular Groove
  Phase Mask Coronagraph},'' {\em \apj}~{\bf 633},  1191--1200 (Nov. 2005).

\bibitem{marois2006}
{Marois}, C., {Lafreni{\`e}re}, D., {Doyon}, R., {Macintosh}, B., and {Nadeau},
  D., ``{Angular Differential Imaging: A Powerful High-Contrast Imaging
  Technique},'' {\em \apj}~{\bf 641},  556--564 (Apr. 2006).

\bibitem{lafreniere2007}
{Lafreni{\`e}re}, D., {Marois}, C., {Doyon}, R., {Nadeau}, D., and {Artigau},
  {\'E}., ``{A New Algorithm for Point-Spread Function Subtraction in
  High-Contrast Imaging: A Demonstration with Angular Differential Imaging},''
  {\em \apj}~{\bf 660},  770--780 (May 2007).

\bibitem{soummer2012}
{Soummer}, R., {Pueyo}, L., and {Larkin}, J., ``{Detection and Characterization
  of Exoplanets and Disks using Projections on Karhunen-Loeve Eigenimages},''
  {\em ArXiv e-prints}~{\bf -} (July 2012).

\bibitem{blanc2003}
{Blanc}, A., {Fusco}, T., {Hartung}, M., {Mugnier}, L.~M., and {Rousset}, G.,
  ``{Calibration of NAOS and CONICA static aberrations. Application of the
  phase diversity technique},'' {\em \aap}~{\bf 399},  373--383 (Feb. 2003).

\bibitem{hartung2003pd}
{Hartung}, M., {Blanc}, A., {Fusco}, T., {Lacombe}, F., {Mugnier}, L.~M.,
  {Rousset}, G., and {Lenzen}, R., ``{Calibration of CONICA static aberrations
  by phase diversity},'' in [{\em Society of Photo-Optical Instrumentation
  Engineers (SPIE) Conference Series}{\nolinebreak\hspace{0.1em}]},  {M.~Iye \&
  A.~F.~M.~Moorwood}, ed., {\em Society of Photo-Optical Instrumentation
  Engineers (SPIE) Conference Series} {\bf 4841},  295--306 (Mar. 2003).

\bibitem{meimon2010}
{Meimon}, S., {Fusco}, T., and {Petit}, C., ``{An optimized calibration
  strategy for high order adaptive optics systems : the Slope-Oriented Hadamard
  Actuation},'' in [{\em Adaptative Optics for Extremely Large
  Telescopes}{\nolinebreak\hspace{0.1em}]},  (2010).

\bibitem{noll76}
Noll, R., ``"zernicke polynomials and atmospheric turbulence",'' {\em Journal
  of the Optical Society of America}~{\bf 66},  207--211 (March 1976).

\bibitem{riaud2012}
{Riaud}, P., {Mawet}, D., and {Magette}, A., ``{Nijboer-Zernike phase retrieval
  for high contrast imaging. Principle, on-sky demonstration with NACO, and
  perspectives in vector vortex coronagraphy},'' {\em \aap}~{\bf 545} (Sept.
  2012).

\bibitem{gendron2003}
{Gendron}, E., {Lacombe}, F., {Rouan}, D., {Charton}, J., {Collin}, C.,
  {Lefort}, B., {Marlot}, C., {Michet}, G., {Nicol}, G., {Pau}, S., {Phan},
  V.~D., {Talureau}, B., {Lizon}, J.-L., and {Hubin}, N.~N., ``{NAOS infrared
  wavefront sensor design and performance},'' in [{\em Society of Photo-Optical
  Instrumentation Engineers (SPIE) Conference
  Series}{\nolinebreak\hspace{0.1em}]},  {Wizinowich}, P.~L. and {Bonaccini},
  D., eds., {\em Society of Photo-Optical Instrumentation Engineers (SPIE)
  Conference Series} {\bf 4839},  195--205 (Feb. 2003).

\bibitem{roberts2004}
{Roberts}, Jr., L.~C., {Perrin}, M.~D., {Marchis}, F., {Sivaramakrishnan}, A.,
  {Makidon}, R.~B., {Christou}, J.~C., {Macintosh}, B.~A., {Poyneer}, L.~A.,
  {van Dam}, M.~A., and {Troy}, M., ``{Is that really your Strehl ratio?},'' in
  [{\em Society of Photo-Optical Instrumentation Engineers (SPIE) Conference
  Series}{\nolinebreak\hspace{0.1em}]},  {Bonaccini Calia}, D., {Ellerbroek},
  B.~L., and {Ragazzoni}, R., eds., {\em Society of Photo-Optical
  Instrumentation Engineers (SPIE) Conference Series} {\bf 5490},  504--515
  (Oct. 2004).

\bibitem{serabyn2007}
{Serabyn}, E., {Wallace}, K., {Troy}, M., {Mennesson}, B., {Haguenauer}, P.,
  {Gappinger}, R., and {Burruss}, R., ``{Extreme Adaptive Optics Imaging with a
  Clear and Well-Corrected Off-Axis Telescope Subaperture},'' {\em \apj}~{\bf
  658},  1386--1391 (Apr. 2007).

\bibitem{quanz2012}
{Quanz}, S.~P., {Crepp}, J.~R., {Janson}, M., {Avenhaus}, H., {Meyer}, M.~R.,
  and {Hillenbrand}, L.~A., ``{Searching for Young Jupiter Analogs around AP
  Col: L-band High-contrast Imaging of the Closest Pre-main-sequence Star},''
  {\em \apj}~{\bf 754},  127 (Aug. 2012).

\bibitem{huelamo2011}
{Hu{\'e}lamo}, N., {Lacour}, S., {Tuthill}, P., {Ireland}, M., {Kraus}, A., and
  {Chauvin}, G., ``{A companion candidate in the gap of the T Chamaeleontis
  transitional disk},'' {\em \aap}~{\bf 528},  L7 (Apr. 2011).

\bibitem{lacour2011_2}
{Lacour}, S., {Tuthill}, P., {Amico}, P., {Ireland}, M., {Ehrenreich}, D.,
  {Huelamo}, N., and {Lagrange}, A.-M., ``{Sparse aperture masking at the VLT.
  I. Faint companion detection limits for the two debris disk stars HD 92945
  and HD 141569},'' {\em \aap}~{\bf 532},  A72 (Aug. 2011).

\bibitem{kasper2010lgs}
{Kasper}, M., {Zins}, G., {Feautrier}, P., {O'Neal}, J., {Michaud}, L.,
  {Rabou}, P., {Stadler}, E., {Charton}, J., {Cumani}, C., {Delboulbe}, A.,
  {Geimer}, C., {Gillet}, G., {Girard}, J., {Huerta}, N., {Kern}, P., {Lizon},
  J., {Lucuix}, C., {Mouillet}, D., {Moulin}, T., {Rochat}, S., and
  {S{\"o}nke}, C., ``{A New Lenslet Array for the NACO Laser Guide Star
  Wavefront Sensor},'' {\em The Messenger}~{\bf 140},  8--9 (June 2010).

\bibitem{otten2012spie}
{Otten}, G.~P., {Kenworthy}, M.~A., and {Codona}, J.~L., ``{Laboratory
  demonstration and characterization of phase-sorting interferometry},'' in
  [{\em {Ground-based and Airborne Instrumentation for Astronomy IV,
  Proceedings of the SPIE}}{\nolinebreak\hspace{0.1em}]},  {McLean}, I.,
  {Ramsay}, S., and {Takami}, H., eds.,  {\bf -} (July 2012).

\end{thebibliography}

\end{document}